\documentclass[draft,grl]{agutex}

\usepackage{lineno}

\usepackage{lscape}
\usepackage{amsmath}
\usepackage{color}
\usepackage[dvips]{graphicx}

\usepackage{array}
\newcolumntype{C}[1]{>{\centering\let\newline\\\arraybackslash\hspace{0pt}}m{#1}}

\setkeys{Gin}{draft=false}
\authorrunninghead{ROUBINET ET AL.}

\titlerunninghead{Streaming potential in fractured rock}

\begin{document}

%
%


\title{Streaming potential modeling in fractured rock: insights into the identification of hydraulically-active fractures}
%
%

%
%



\authors{D. Roubinet\altaffilmark{1,3},
N. Linde\altaffilmark{1}, D. Jougnot\altaffilmark{2}, 
and J. Irving\altaffilmark{1}}
\altaffiltext{1}{Applied and Environmental Geophysics Group, Institute of Earth Sciences, University of Lausanne, Switzerland.}

\altaffiltext{2}{Sorbonne Universit\'es, UPMC Universit\'e Paris 06, CNRS, EPHE, UMR 7619 METIS, Paris, France}

\altaffiltext{3}{Corresponding author: delphine.roubinet@unil.ch}
%
%

%
%



%
%



%
%

%

\begin{article}

This paper has been accepted for publication in Geophysical Research Letters: \\
\textit{Roubinet, D., N. Linde, D. Jougnot, and J. Irving (2016), Streaming potential modeling in fractured rock: Insights into the identification of hydraulically active fractures, Geophys. Res. Lett., 43, doi:10.1002/2016GL068669.}

\newpage

\section*{Keypoints}
\begin{itemize}
\item Highly efficient discrete-dual-porosity approach for simulating self-potential (SP) signals in fractured porous media
\item Determinant role of matrix fluid flow in the generation of fractured-rock SP signals
\item Hydraulically-active fractures with significant fracture-matrix interactions can be identified with the SP method
\end{itemize}

\section*{Abstract}
Numerous field experiments suggest that the self-potential (SP) geophysical method may allow for the detection of hydraulically-active fractures and provide information about fracture properties.  However, a lack of suitable numerical tools for modeling streaming potentials in fractured media prevents quantitative interpretation and limits our understanding of how the SP method can be used in this regard.  To address this issue, we present a highly efficient two-dimensional discrete-dual-porosity approach for solving the fluid flow and associated self-potential problems in fractured rock.  Our approach is specifically designed for complex fracture networks that cannot be investigated using standard numerical methods. We then simulate SP signals associated with pumping conditions for a number of examples to show that (i) accounting for matrix fluid flow is essential for accurate SP modeling, and (ii) the sensitivity of SP to hydraulically-active fractures is intimately linked with fracture-matrix fluid interactions.  This implies that fractures associated with strong SP amplitudes are likely to be hydraulically conductive, attracting fluid flow from the surrounding matrix.
%
%

\section{Introduction}
Quantification of fluid flow in fractured media is an outstanding challenge that is critically important in a wide variety of research fields and applications including hydrogeology, geothermal energy, hydrocarbon extraction, and the long-term storage of toxic waste \citep[e.g.,][]{Carneiro2009,Kolditz1998,Rotter2008}.  In the context of resource extraction, the presence of fractures is generally considered to be an advantage as they facilitate access to materials stored in the matrix.  Conversely, with regard to the storage of toxic elements, fractures represent a risk of leakage and subsequent migration of pollutants deep into the subsurface.  In all cases, the detection of fractures and the characterization of their properties in natural environments is a required and critical task \citep{NAP2015}.

Geophysics offers a variety of tools that can provide important information on subsurface structure, physical properties, and fluid flow in a non-invasive manner \citep[e.g.,][]{Hubbard2011}.  Most geophysical techniques infer fluid flow by data or model differencing in time or space; that is, they are not directly sensitive to flow occurring at the time of the measurements. An exception is the self-potential (SP) method, which is of particular interest for hydrogeological applications because of its direct sensitivity to water flowing in the subsurface \cite[i.e., the streaming potential;][]{revil2013self}. This phenomenon is intimately linked to the presence of an excess charge in the pore water that counter-balances electric charges at the mineral-pore water interface. When water flows through the pore, it gives rise to a streaming current and an associated streaming potential. The direct sensitivity to subsurface fluid flow makes the SP method particularly interesting for the study of fractured rocks, in which flow is often highly channelized in a small fraction of the rock volume \cite[e.g.][]{Berkowitz2002}.

Previous studies have demonstrated the ability of the SP method to detect groundwater flow in fractured media \cite[e.g.,][]{fagerlund2003detecting,Maineult2013}, determine the orientation of hydraulically-active fractures through azimuthal measurements at the ground surface \citep{wishart2006self,wishart2008fracture}, and localize water leakage through a single fracture \citep{revil2015passive}.  Although such studies clearly demonstrate the potential utility of SP measurements in fractured-rock investigations, there is currently a dearth of numerical modeling tools to simulate SP responses in fractured media, the latter of which are required for quantitative interpretation and inversion of field data.  Indeed, whereas fully discretized finite-element approaches are regularly used to simulate SP signals in porous media \cite[e.g.][]{revil2013self}, such methods quickly become computationally prohibitive when considering fractured rock where the fractures must be discretized at a fine spatial scale.

In this paper, we address the above challenges and present a highly efficient, discrete-dual-porosity (DDP) approach for simulating fluid flow and streaming potentials in fractured porous media.  Our approach builds on the 2D electric-current-flow model developed by \cite{Roubinet2014}, and importantly considers the exchange of water between fractures and the surrounding matrix.  The proposed modeling approach is specifically designed for highly heterogeneous fractured porous media that cannot be handled by standard numerical methods. Indeed, simulations are found to be approximately 50 times faster than standard finite-element methods for simple configurations for which a comparison can be made.  We use our numerical approach to simulate field-scale SP experiments under pumping conditions in order to demonstrate that (i) considering the fluid flowing in the matrix is absolutely essential for accurate simulation of SP signals in fractured rock; (ii) strong SP signals are observed for hydraulically-active fractures having significant fracture-matrix fluid interactions; and (iii) the detection of individual hydraulically-active fractures by SP measurements is feasible only when the hydrogeological response is determined by a few dominant fractures.

\section{Methodological background}\label{sec:method}

\subsection{Governing equations}\label{sec:equations}
Under steady-state conditions, the Darcy velocity $\mathbf u$ [m/s] of an incompressible fluid can be described by Darcy's law and Darcy-scale mass conservation \citep[e.g.,][]{Bear2010},
\begin{linenomath*}
\begin{equation}\label{eq:fluid_flow} 
\mathbf u = - K \nabla h, \quad \nabla \cdot \mathbf u = 0,
\end{equation}
\end{linenomath*}
where $K$ [m/s] and $h$ [m] are the hydraulic conductivity and hydraulic head, respectively.  Here, source or sink terms are not considered, and the hydraulic conductivity is defined as $K = \kappa \rho g /\mu$ with $\kappa$ [m$^2$] the medium permeability, $\rho$ [kg/m$^3$] the fluid density, $g$ [m/s$^2$] the gravitational acceleration, and $\mu$ [kg/(m$\cdot$s)] the fluid dynamic viscosity. 

Considering the presence of an electrical double layer that results in an excess of charge in the porewater, the water flowing through the medium drags a part of this excess charge $\bar Q_v^{eff}$ [C/m$^3$].  This generates an electrokinetic source current density $\mathbf J_s$ [A/m$^2$], which can be defined as \citep{Titov2002}
\begin{linenomath*}
\begin{equation}\label{eq:J_s} 
\mathbf J_s = \bar Q_v^{eff}  \mathbf u.
\end{equation}
\end{linenomath*}
In the quasi-static limit, the impact of this streaming current on the electric potential is described by the following charge conservation equation \citep{Sill1983}:
\begin{linenomath*}
\begin{equation}\label{eq:cons_elec} 
\nabla \cdot \left(-\sigma \nabla \varphi + \mathbf J_s\right) = 0,
\end{equation}
\end{linenomath*}
where $\sigma$ [S/m] and $\varphi$ [V] are the bulk electrical conductivity and electric potential, respectively.

\subsection{Overall modeling strategy}\label{sec:model_strategy}
We wish to solve equations~\eqref{eq:fluid_flow}-\eqref{eq:cons_elec} in complex fractured rock formations that are characterized by a large contrast between the permeability of the fractures and that of the surrounding matrix.  For this purpose, we consider a DDP representation in which the fractures are explicitly represented.  In accordance with existing DDP formulations developed for fluid flow in fractured reservoirs \citep[e.g.,][]{Lee2001,Li2008}, each fracture is represented using two parallel plates between which the flow is assumed to be laminar.  This means that (i) the fracture permeability can be defined as $\kappa_f=(b_f)^2/12$ where $b_f$ is the fracture aperture \citep[e.g.,][]{Snow1969}; and (ii) the excess charge can be evaluated numerically by adapting the strategy of \cite{jougnot2012derivation} for a single capillary tube to the case of two infinite plates having known separation.

Based on this DDP representation, we solve successively equations~\eqref{eq:fluid_flow} and \eqref{eq:cons_elec} in two dimensions using the hydrogeological and self-potential model formulations described in Sections~\ref{sec:model_hydro} and \ref{sec:model_elec}, respectively.  In each case, the fractures and matrix are divided into fracture segments and matrix blocks having constant properties, which are coupled through an exchange coefficient defined at the matrix-block scale.  Please see \cite{Roubinet2014} for a detailed description of the representation and discretization methods used to model the geological structures.

\subsection{Hydrogeological model formulation}\label{sec:model_hydro}
As the fluid flow problem in equation~\eqref{eq:fluid_flow} is mathematically equivalent to the electric current flow problem in equation~\eqref{eq:cons_elec} with $\mathbf J_s$ set to zero and with the hydraulic conductivity and hydraulic head replacing the electrical conductivity and electric potential, the DDP approach described by \cite{Roubinet2014} for electric current flow can be employed to determine the distribution of hydraulic head in the fractures and matrix.  This distribution is then used to evaluate the Darcy velocity of the fluid circulating in these structures, as well as the Darcy velocity of the fluid exchanged between them.  The latter, expressed in the direction perpendicular to each considered fracture, is given by
\begin{linenomath*}
\begin{equation}\label{eq:ufm} 
u_{fm} = -\frac{K_m (h_m-h_f)}{<d>},
\end{equation}
\end{linenomath*}
where $K_m$ is the matrix hydraulic conductivity, $h_m$ and $h_f$ are the matrix and fracture hydraulic heads, respectively, and $<d>$ is the average normal distance between the fracture and each point of the surrounding matrix block \citep{Lee2001,Li2008}.  Whereas expression~\eqref{eq:ufm} is usually defined at the matrix-block scale, note that we define it at the fracture-segment scale with $h_f$ the hydraulic head averaged along the considered fracture segment.

\subsection{Self-potential model formulation}\label{sec:model_elec}
The presence of the previously described fluid flows combined with an excess of charge in the porewater implies the generation of source current densities in the fractures and matrix, which we denote by $\mathbf J_{s,f}$ and $\mathbf J_{s,m}$, respectively.  We now describe how we account for these current densities in the electrical problem at the fracture, fracture-network, and matrix-block scales, with particular focus on the additional terms required in comparison with the DDP approach described by \cite{Roubinet2014}.

At the fracture scale, consider the charge conservation equation~\eqref{eq:cons_elec} in two dimensions with a constant fracture electrical conductivity $\sigma_f$:
\begin{linenomath*}
\begin{equation}\label{eq:cons_elec_seg2D} 
-\sigma_f \left(\partial_{x_f}^2 \varphi_f + \partial_{y_f}^2 \varphi_f \right) = -\partial_{x_f} J_{s,f}^{x_f} - \partial_{y_f} J_{s,f}^{y_f},
\end{equation}
\end{linenomath*}
where $x_f$ and $y_f$ denote the spatial variables parallel and perpendicular to the considered fracture, respectively, $\varphi_f$ denotes the electric potential in this fracture, and $J_{s,f}^{x_f}$ and $J_{s,f}^{y_f}$ are the components of $\mathbf J_{s,f}$ in the $x_f$ and $y_f$ directions, respectively.  As our 2D DDP formulation is based on a 1D representation of fractures, we average equation~\eqref{eq:cons_elec_seg2D} over the aperture of the considered fracture.  Assuming that the variation of $J_{s,f}^{x_f}$ along the fracture is negligible ($\partial_{x_f} J_{s,f}^{x_f}=0$~A/m$^3$), this yields
\begin{linenomath*}
\begin{equation}\label{eq:cons_elec_seg1D} 
-\sigma_f \partial_{x_f}^2 \bar \varphi_f = -\mathcal Q_{fm}^{E}-\mathcal Q_{fm}^{SP},
\end{equation}
\end{linenomath*}
where $\bar \varphi_f = \frac{1}{b_f} \int_0^{b_f} \varphi_f(x_f,y_f) \mathrm d y_f$ is the electric potential averaged over the fracture aperture, and 
$\mathcal Q_{fm}^{E} = -\frac{\sigma_f}{b_f} \left[\partial_{y_f} \varphi_{f(y_f=b_f)}-\partial_{y_f} \varphi_{f(y_f=0)} \right]$ 
and 
$\mathcal Q_{fm}^{SP} = \frac{1}{b_f} \left[ J_{s,f(y_f=b_f)}^{y_f}-J_{s,f(y_f=0)}^{y_f} \right]$
are electric source/sink terms representing exchange occurring at the fracture-matrix interfaces related to the variations of electric potential and hydraulic head, respectively.  In our dual-porosity formulation, this leads to $\mathcal Q_{fm}^{E}$ being expressed as a function of the difference in electric potential between the fracture and matrix \citep{Roubinet2014} and to $\mathcal Q_{fm}^{SP}$ being given, using expression~\eqref{eq:J_s}, as
\begin{linenomath*}
\begin{equation} 
\mathcal Q_{fm}^{SP} = \bar Q_{v,f}^{eff} u_{fm}/b_f,
\end{equation}
\end{linenomath*}
where $\bar Q_{v,f}^{eff}$ is the effective excess charge in the considered fracture and $u_{fm}$ is the Darcy velocity of the fluid going from this fracture to the surrounding matrix (Section~\ref{sec:model_hydro}).  

At the fracture-network scale, charge conservation at each fracture intersection is enforced by integrating equation~\eqref{eq:cons_elec} over the considered intersection.  Using Gauss' Divergence Theorem with $C_i$ denoting the contour of the intersection and $\vec n_{C_i}$ its outward unit normal vector leads to the expression
\begin{linenomath*}
\begin{equation}\label{eq:cons_fract_inter}
- \int_{C_i} \sigma_f \nabla \varphi_f \cdot \vec n_{C_i} \mathrm d {C_i} = 
- \int_{C_i} \bar Q_{v,f}^{eff}  \mathbf u_f \cdot \vec n_{C_i} \mathrm d {C_i},
\end{equation}
\end{linenomath*}
where $\mathbf u_f $ is the Darcy velocity of the fluid in the fracture.  As each fracture intersection is the shared extremity of several fracture segments, expression~\eqref{eq:cons_fract_inter} is discretized as the sum of the integrals over the apertures of these fracture segments.  The left-hand side is obtained using the analytical expression of equation~\eqref{eq:cons_elec_seg1D} and the right-hand side is taken into account as a source term.

Finally, in the matrix, charge conservation is again enforced by integrating equation~\eqref{eq:cons_elec} over each matrix block having volume $V_m$ and applying Gauss' Divergence Theorem.  Using $C_m$ to denote the contour of a matrix block and $\vec n_{C_m}$ its outward unit normal vector leads to
\begin{linenomath*}
\begin{align}\label{eq:cons_matrix_block}
& -  \int_{C_m} \sigma_m \nabla \varphi_m \cdot \vec n_{C_m} \mathrm d C_m = \\ \nonumber 
& - \int_{C_m} \bar Q_{v,m}^{eff}  \mathbf u_m \cdot \vec n_{C_m} \mathrm d {C_m}   
 + \int_{V_m} \left(\mathcal Q_{fm}^E+\mathcal Q_{fm}^{SP} \right) \mathrm dV_m,
\end{align}
\end{linenomath*}
where the first term on the right-hand side is related to the electrokinetic source current density $\mathbf J_{s,m}$ exchanged between the matrix blocks, and the second term is related to fracture-matrix exchanges occurring inside each block.
We discretize expression~\eqref{eq:cons_matrix_block} using a modified finite-volume method where the terms related to the presence of fluid flow are taken into account as source terms.


\section{Method validation}\label{sec:validation}
\begin{figure}
\centering
\includegraphics[width=15cm]{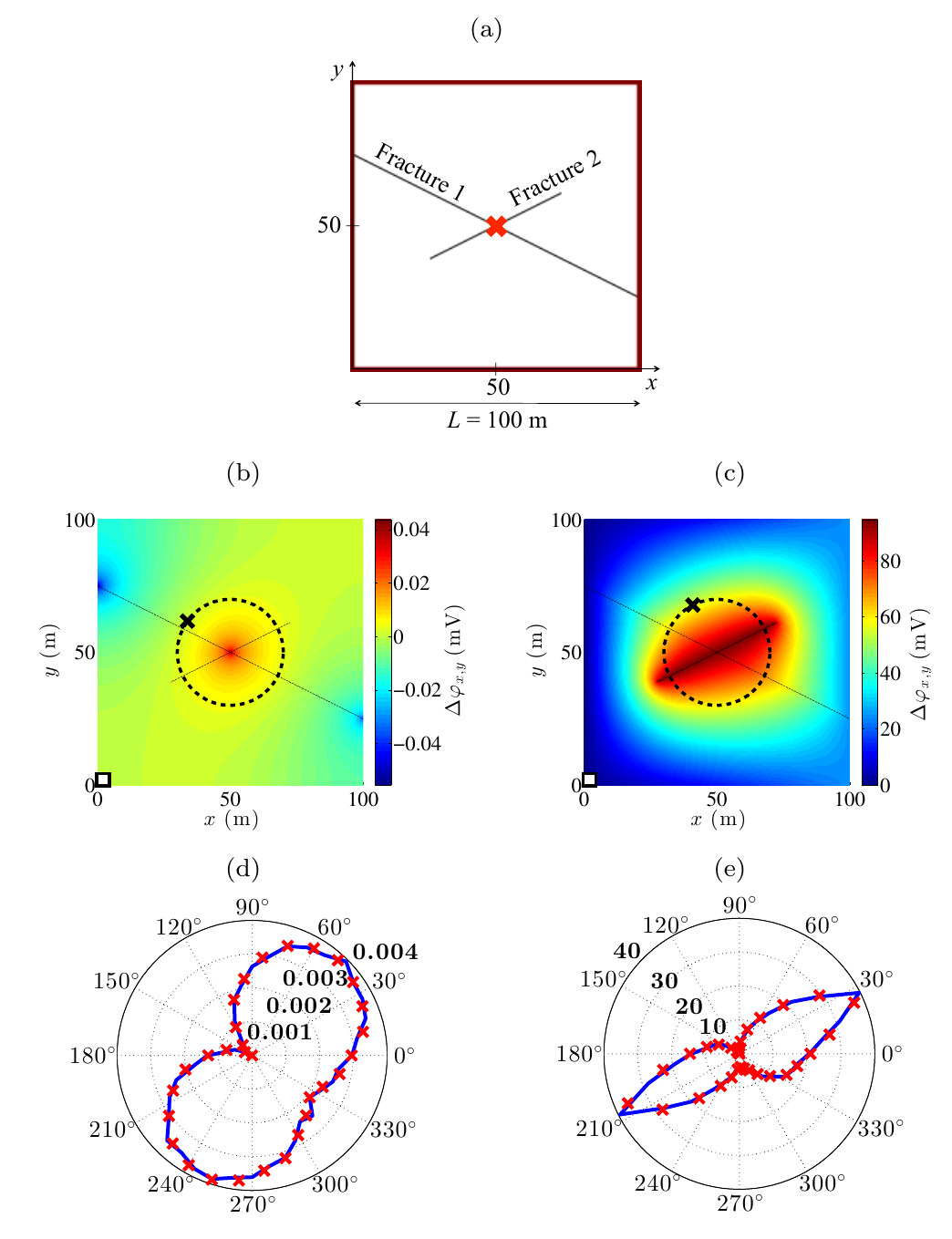}
\caption{(a) Configuration used to validate our modeling approach where the red cross represents the position of the considered pumping well.  (b-c) Spatial distribution of the SP signal $\Delta \varphi_{x,y}$ (in mV) computed with our DDP approach where the white square represents the position of the reference electrode.  (d-e) Polar plots of the SP signal $\Delta \varphi_r^*$ (in mV) with respect to a reference located at the black cross in (b) and (c), respectively, and computed with our DDP approach (blue lines) and a finite element solution (red crosses).
Note that matrix fluid flow was ignored in (b) and (d) and accounted for in (c) and (e). }
\label{fig:validation}
\end{figure}

To validate our modeling approach, we consider the fracture, matrix, and fluid properties in Table~\ref{materialprop} and the basic 2D fracture configuration shown in Figure~\ref{fig:validation}a.  In this configuration, two fractures of infinite vertical extent intersect at position $(x,y) = (50,50)$~m. The fractures are oriented at angles of $-26.6^\circ$ (Fracture~1) and $26.6^\circ$ (Fracture~2), and have widths (in the $x$-$y$ plane) of $111.8$~m (Fracture~1) and $50$~m (Fracture~2).  The fracture permeability and effective excess charge were evaluated as described in Section~\ref{sec:model_strategy} and the fracture electrical conductivity was set equal to the water conductivity $\sigma_w = 5 \times 10^{-2}$~S/m as we consider clean fractures (i.e., no filling).  For the matrix, the permeability was chosen to be representative of sandstone \citep{Schon2011} and the excess charge was evaluated from the permeability using the empirical relationship defined by \cite{Jardani2007}.  The matrix electrical conductivity was determined using Archie's law: $\sigma = \sigma_w \phi^{m}$, where $\phi$ is the porosity and $m$ the cementation exponent (set equal to 2).

For the fluid flow problem, we consider: (i) Dirichlet hydraulic head boundary conditions equal to $1$~m and $0$~m on the left and right sides of the domain, respectively; (ii) Dirichlet boundary conditions varying linearly between these two values along the top and bottom; and (iii) a sink term applied at the center of the domain corresponding to pumping at a rate of $0.8 \times 10^{-3}$ m$^2$/s.  To study the impact of water flowing in the low-permeability matrix on the generated SP signal, we solve the fluid flow problem first only in the fractures (i.e., ignoring matrix flow) and then over the whole domain. 
For the streaming potential problem, we consider: (i) a current insulation condition on all borders of the domain; and (ii) a reference electrode located at $(x,y)=(0,0)$~m with a specified potential of $0$~V.  

Figures~\ref{fig:validation}b and c show the distribution of the electric potential difference $\Delta \varphi_{x,y}(x,y)=\varphi_{x,y}(x,y)-\varphi_{x,y}(0,0)$ for the cases where matrix fluid flow was ignored and accounted for, respectively.  These results were computed using our DDP approach with 201 matrix blocks in each direction, as well as with the COMSOL Multiphysics 4.3 finite-element software package.  In the latter case, using the ``extremely coarse'' meshing option resulted in $8.1 \times 10^{5}$ model elements and a combined meshing/computation time of $1.5$~hours on a 2.9~GHz laptop computer with 8~GB RAM. In comparison, our DDP approach required only $4.1  \times 10^{4}$ model elements and the total computation time was less than 2 minutes.  The mean absolute difference between the results of our DDP approach and those of the finite-element solution is $2.6 \times 10^{-5}$~mV for the case with no matrix flow and $3.2 \times 10^{-1}$~mV when it was considered, which demonstrates a good agreement between the two modeling methods.  This agreement is also observed in Figures~\ref{fig:validation}d and e, which contain polar plots of the potential difference in Figures~\ref{fig:validation}b and c along the dashed circle having a radius of $20$~m.  Here the plotted values $\Delta \varphi_r^*(\theta)$ were calculated relative to the minimum along the circle $\Delta \varphi_r^{min}$, whose position is shown by a black cross in Figures~\ref{fig:validation}b and c. That is,
\begin{linenomath*}
\begin{align}\label{eq:phi_r_star}
\Delta \varphi_r^*(\theta)=\Delta \varphi_r(\theta)-\Delta \varphi_r^{min}.
\end{align}
\end{linenomath*}

The results of the validation (Figure~\ref{fig:validation}) provide important insight into whether it is reasonable to neglect matrix fluid flow when modeling streaming potentials in fractured media. As seen in Figures~\ref{fig:validation}b and d, making this assumption results in extremely small SP values compared to the case where matrix flow and matrix-fracture interactions are taken into account (Figures~\ref{fig:validation}c and e). This is despite the fact that, in our example, the matrix permeability was set to be 8 orders of magnitude smaller than that of the fractures and thus the matrix can be considered as impermeable from a flow modeling perspective. Indeed, whereas extremely small matrix permeabilities are generally ignored for flow computations in fractured media, they cannot be neglected when modeling the SP response. These results also imply that if the matrix material is strictly impervious, or if the fracture network is so well connected that fracture-matrix fluid exchange is minimal, then the associated SP signals will be negligible.
 
The results obtained for the case where matrix fluid flow was accounted for (Figures~\ref{fig:validation}c and e) also provide information concerning the sensitivity of the SP method to fracture-matrix fluid exchanges.  As Fracture~2 does not intersect the domain boundaries (Figure~\ref{fig:validation}a), the fluid circulating in this fracture during the pumping experiment is provided by the surrounding matrix.  Conversely, the fluid circulating in Fracture~1 is mostly provided by the Dirichlet conditions enforced at the extremities of the modeling domain.  That is, Fracture~2 is characterized by strong fracture-matrix fluid interactions which result in (i) strong variations of the SP response $\Delta \varphi_{x,y}$ perpendicular to this fracture, and (ii) small variations of $\Delta \varphi_{x,y}$ along this fracture where the maximum value of $\Delta \varphi_{x,y}$ is observed (Figure~\ref{fig:validation}c).  As a result, the maximum value of $\Delta \varphi_{r}^*$ is observed in the direction of Fracture~2 (Figure~\ref{fig:validation}e), which demonstrates that azimuthal SP measurements would be more sensitive to Fracture~2 than Fracture~1.

%

\section{Results for complex fractured media}\label{sec:results}
\begin{figure}[h]
\centering
\includegraphics[width=15.3cm]{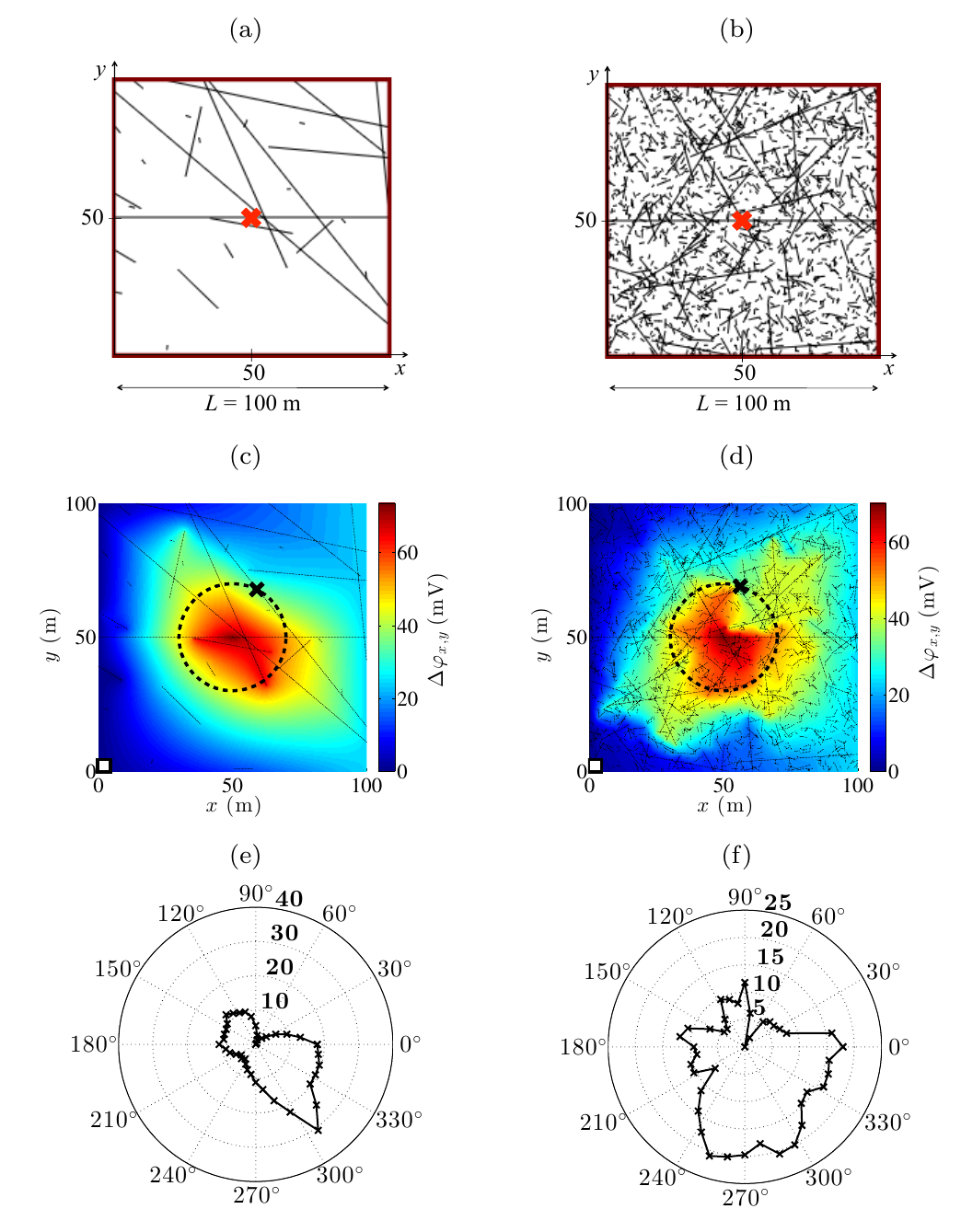}
\caption{(a-b) Studied fractured domains where the red cross represents the position of the considered pumping well. (c-d) Corresponding spatial distribution of the SP signal $\Delta \varphi_{x,y}$ (in mV) computed with our DDP approach where the white square represents the position of the reference electrode. (e-f) Polar plots of the SP signal $\Delta \varphi_r^*$ (in mV) with respect to the black crosses located in (c) and (d), respectively, and computed with our DDP approach.}
\label{fig:results}
\end{figure}

We now use our modeling approach to investigate how the SP method may help to identify hydraulically-active fractures in complex fractured media.  To this end, we consider the two 2D fracture configurations shown in Figures~\ref{fig:results}a and b.  These $100 \times 100$~m regions contain (i) a primary fracture at position $y=50$~m extending from $x=0$~m to $100$~m that is characterized by an aperture of $2 \times  10^{-3}$~m, and (ii) a random distribution of fractures all having an aperture of $10^{-3}$~m.  In the latter case, the fracture positions and orientations were drawn from a uniform distribution, whereas a power-law distribution was considered for their lengths using power law exponents equal to $1.5$ (Figure~\ref{fig:results}a) and $2.5$ (Figure~\ref{fig:results}b) and percolation parameter equal to $6$.  Justifications and examples of fracture networks corresponding to these distributions and parameters can be found in \cite{Bonnet2001}, \cite{Bour1997}, and \cite{Roubinet2010a}. 

We consider the same fracture, matrix, and fluid properties as before (Table~\ref{materialprop}), as well as the same boundary and source conditions for the fluid flow and streaming potential problems with, again, pumping applied at the center of the domain.  Note that matrix fluid flow and fracture-matrix fluid exchange are taken into account in these examples.  Figures~\ref{fig:results}c-d show the distribution of the SP response $\Delta \varphi_{x,y}$ computed with our modeling approach using 201 matrix blocks in each direction, and Figures~\ref{fig:results}e-f show the corresponding polar plots of $\Delta \varphi_r^*$ defined in equation~\eqref{eq:phi_r_star} with, again, the position of $\Delta \varphi_r^{min}$ represented with a black cross in Figures~\ref{fig:results}c-d.

For the fractured medium in Figure~\ref{fig:results}a, we observe that the maximum value of $\Delta \varphi_r^*$ is $31$~mV and that this value is obtained when $\theta=306^\circ$ (Figure~\ref{fig:results}e).  This corresponds to the location of a dead-end fracture that is hydraulically connected and close to the pumping well, which implies that the considered fracture contributes significantly to the pumped volume of fluid and that this contribution is mostly provided by the surrounding matrix.  In other words, fluid fracture-matrix exchanges are maximized at this position, which results in a strong SP signal.  Figures~\ref{fig:results}c and e also show that $\Delta \varphi_r^*$ ranges from $10$~mV to $20$~mV for angles corresponding to (i) the primary fracture in which the pumping is applied; and (ii) fractures that are connected to the domain boundaries and hydraulically connected to the pumping well.  Conversely, $\Delta \varphi^*_r$ is smaller than $10$~mV for angles associated with regions where there is no fracture that is well connected to the pumping well.  This demonstrates that the contrast in hydraulic and electrical properties between fractures and matrix, as well as the corresponding fracture-matrix exchanges, result in variations of $\Delta \varphi_{x,y}$ that are smaller along the hydraulically-active fractures than perpendicular to these fractures.  

For the fractured medium presented in Figure~\ref{fig:results}b, a different behavior for $\Delta \varphi_r^*$ is observed (Figure~\ref{fig:results}f).  In this case, we do not see any localized large values characteristic of dominant hydraulically-active fractures having important fracture-matrix interactions.  We observe that (i) the largest values of $\Delta \varphi_r^*$ around $20$~mV are obtained when $\theta$ ranges from $252^\circ$ to $297^\circ$; and (ii) the smallest values of $\Delta \varphi_r^*$, defined here as $\Delta \varphi_r^* \leq 5$~mV, are obtained when $63^\circ \leq \theta \leq 72^\circ$ and $135^\circ \leq \theta \leq 144^\circ$.  The latter regions having small SP values correspond to regions containing comparatively small fluid flow (not shown).

Note that the simulations presented above represent scenarios where (i) a natural hydraulic gradient is locally perturbed by pumping, and (ii) the pumping rate is large enough such that the simulated azimuthal electrical measurements are not significantly affected by the boundary conditions.  Further analysis could help to assess the impact of these boundary conditions and their distance from the pumping well on the simulated SP signals.  This could be easily done, for example, by comparing the signals obtained for domains of increasing size associated with consistent boundary conditions and fracture network properties.


\section{Conclusions}\label{sec:conclusions}
We have presented a highly efficient and accurate 2D discrete dual-porosity approach for modeling streaming potentials in complex fractured porous media. A key feature of this approach is that it accounts for fluid and electric current flow in and between the fractures and surrounding matrix. For a simple configuration with two intersecting fractures and pumping, we found that our method is 50 times faster and provides the same response as a fully discretized finite-element numerical solution. We also saw in this example that ignoring matrix-fracture interactions may lead to simulated SP responses that are not only wrong in overall appearance, but also orders of magnitude too small in amplitude. These results and further examples based on more complex fractured domains demonstrate that the SP method, when applied to fractured media, is primarily sensitive to hydraulically-active fractures having important fracture-matrix exchange. The latter finding opens up exciting possibilities to remotely and non-invasively identify fracture-matrix exchange using SP measurements. It would also be straightforward to include our new modeling approach within an inversion framework to quantify these exchanges from field data. For more densely fractured media, we approach an upscaled effective medium response, in which it becomes impossible to identify individual fractures. However, in this case, strong SP signals will still inform us about regions that are (in an average sense) well connected to the pumping well, whereas small SP signals will indicate regions that are poorly connected to the pumping well.  Future work will focus on simulating SP signals in 3D fractured domains.  For this purpose, the required numerical method will build on either 2.5D or 3D formulations of our DDP approach, which are currently in development for modeling electrical current flow in fractured porous media.

\section*{Ackowledgements}
The authors wish to thank Prof. Andr\'e Revil for insightful comments that greatly helped to improve the quality of this work.  We are also grateful for the constructive comments offered by two anonymous reviewers.

\end{article}
%
%

%
\begin{table}
\caption{Properties used for the numerical validation (Figure~\ref{fig:validation}).}
\begin{center}
\begin{tabular}{lC{3cm}C{3cm}C{3.5cm}}
	\hline
	& Fracture~1 & Fracture~2 & Matrix \\
	\hline
	$\phi$ [-] & 1 & 1 & 0.1 \\
	$b$ [m] & $2 \times 10^{-3}$ & $10^{-3}$ & - \\
	$\kappa$ [m$^2$] & $3.3 \times 10^{-7}$ & $8.3 \times 10^{-8}$ & $10^{-15}$ \\
	$\sigma$ [S/m] & $5 \times 10^{-2}$ & $5 \times 10^{-2}$ & $5 \times 10^{-4}$ \\
	$\bar{Q}_v^{eff}$ [C/m$^3$] & $3.6 \times 10^{-5}$ & $1.4 \times 10^{-4}$ & $1.2 \times 10^{3}$ \\
	\hline \hline
	$\rho = 10^3$~kg/m$^3$ & \multicolumn{2}{c}{$g=9.8$~m/s$^2$} & $\mu=10^{-3}$~kg/(m$\cdot$s)\\
	\hline
\end{tabular}
\label{materialprop} 
\end{center}
\end{table}

%
%
%
%
%
%


\end{document}